\documentclass[prl,a4paper,superscriptaddress,twocolumn,showpacs,amsmath,amssymb,floatfix,amstex]{revtex4}%
\newcommand{\ket}[1]{| #1\rangle}                       %
\newcommand{\brac}[1]{\langle #1|}                      %
\newcommand{\su}[1]{\text{\textbf{\textsf{#1}}}}        %
\newcommand{\s}{\su{S}}                                 %
\newcommand{\se}{\s_\epsilon}
\newcommand{\cpq}{c_{\epsilon}(p,q)}                    %
\newfont{\Bb}{msbm10}                                   %
\newcommand{\Tqp}{T_{qp}}           									%
\newcommand{\Tqpd}{T_{qp}^{\dagger}}     						%
\newcommand{\equa}[1]{Eq.~(\ref{#1})}       						%
\usepackage{graphicx,graphics,wrapfig,rotating}	% Include figure files
\usepackage{dcolumn}
\usepackage{bm,pstricks}
\usepackage{hyperref}

\newrgbcolor{darkblue}{0.165 0.1 0.43}
\newrgbcolor{lcolor}{0. 0. 0.80}
\newrgbcolor{gcolor2}{.80 .80 1.}
\newrgbcolor{terra}{0.73 0.28 0.22}
\newrgbcolor{qcggreen}{0.71 0.78 0.68}
\newrgbcolor{qcgyellow}{1. 0.85 0.5}
\newrgbcolor{darkgreen}{0.25 0.46 0.22}
\begin{document}
%\preprint{APS/123-QED}
\title{Entanglement-screening by nonlinear resonances} 
\author{Ignacio Garc\'{\i}a-Mata}
\email{garcia@irsamc.ups-tlse.fr}
\affiliation{%
\mbox{Laboratoire de Physique Th\'eorique, UMR 5152 du CNRS,
Universit\'e Paul Sabatier, 31062 Toulouse Cedex 4, France}
}
\author{Andr\'e R. R. Carvalho}
\affiliation{%
Department of Physics,
Faculty of Science,
Australian National University
ACT 0200, Australia
}%
\author{Florian Mintert}
\affiliation{%
Physics Department, Harvard University, Cambridge, Massachusetts 02138, U.S.A.}
\date{\today}% 
%%%
\author{Andreas Buchleitner}
\affiliation{%
\mbox{Max-Planck-Institut f\"ur Physik komplexer Systeme, N\"othnitzer  
Str.~38, D-01187
Dresden, Germany.}}
\begin{abstract}
We show that nonlinear resonances in a classically mixed phase space allow to define
generic, strongly entangled multi-partite quantum states.   
The robustness of their multipartite entanglement 
increases with the particle number, i.e. in the semiclassical limit, for those
classes of diffusive 
noise which assist the quantum-classical transition.
\end{abstract}
\maketitle

The ultimate success of the quantum information and computation program will
depend on our theoretical understanding and experimental control of quantum
entanglement. In order to compete with the available classical supercomputing
resources, a future quantum computer will have to be composed of a large,
i.e., at least mesoscopic number of qubits, and their coherence will need to
be preserved over a significant period of time. Contemplating the fact that
entanglement is a manifestation of multi-particle coherence, and that the
density of states explodes exponentially with the particle number, it is easy
to appreciate the dimension of the challenge ahead. 

So far, little is known about entanglement in open quantum systems -- where
``open'' refers to the unavoidable coupling to uncontrolled degrees of freedom
in the ``environment'', which eventually induces decoherence in the system
dynamics. Only recently 
\cite{PhysRevA.65.012101,MCKB,fine:153105,carvalho-2005,yu:140403,petitjean2006,puentes-2006} 
have there been first theoretical and experimental
attempts to characterize entanglement dynamics under decoherence, but a
sufficiently general picture still has to emerge. In particular, most studies
did so far focus on specific classes of highly entangled
W, GHZ, or cluster states, and on their specific robustness against
certain sources of decoherence \cite{arrc,dur:180403}. As the number of particles increases, the
faithful experimental generation and probing of these states tends to become
more and more difficult, with rapidly increasing experimental
overhead. Furthermore, whether and in which sense their entanglement
properties can be considered as ``generic'' is a largely open issue, given the
complicated topology of state space.

In our present contribution, we will adopt a different perspective, which
imports some generic features of quantum dynamics with underlying mixed,
regular-chaotic phase space structure,
i.e., from quantum chaos \cite{javier2006}. 
In contrast to earlier studies of the impact of mixed phase space dynamics
\cite{nemes1998,miller1999,fujisaki2003} and nonlinear light-matter
interaction \cite{kowalewska2006,olsen2006} on bipartite entanglement, we are
here interested in the multipartite 
limit of large particle numbers. This is of crucial importance in the
context of entanglement scaling alluded to above, and will be identified with the
semiclassical 
limit of progressively finer (quantum) resolution of classical phase space
structures, by a suitable definition of many-particle basis states in terms of
classical phase space coordinates. We will see that nonlinear resonances,
which are ubiquitous in classical Hamiltonian systems \cite{licht}, naturally define
strongly -- though not maximally -- entangled multipartite quantum
states. This non-optimality is compensated by the nonlinear
resonance structure providing a natural shelter against certain types of
decoherence: Indeed, the robustness of
the associated multipartite entangled states is found to {\em increase} with the number of
particles, i.e., in the semiclassical limit.

We start out with a system of $k$ qubits that lives on a 
%is described by a 
Hilbert space with tensor structure 
${\cal H}={\cal H}_1\otimes {\cal H}_2\otimes\ldots\otimes {\cal H}_k$, 
where each factor space ${\cal H}_j$ has dimension two. The ``computational
basis''  $\{\ket{i}\}$, which spans $\cal H$, is given by the $k$-particle
product states (i.e., by binary $k$-strings, which we identify with the
binary representations of $i=0\ldots N$), and has dimension $N=2^k$.   
Maps on classical phase space can be implemented efficiently in such
systems after suitable identification of the computational with the position
basis $\{\ket{q_i}\}$, upon the substitution $q_i=i/N$ \cite{miquel,ronc,scott}.
Then, the corresponding momentum basis $\{\ket{p_{i}}\}$ is given by the discrete Fourier
transform of $\{\ket{q_i}\}$ \cite{hannay}. Thus, the phase space is wrapped on a
torus, with a phase space area $1/N$ occupied by a
single position basis state. 
Coherent (or minimum uncertainty) states are obtained as Gaussian wave packets
of width $1/\sqrt{N}$ both in position and momentum \cite{nonn2}. 
Accordingly, the effective Planck constant is
given by $\hbar_{\rm eff}=1/2\pi N$, and the semiclassical limit 
$\hbar_{\rm eff}\rightarrow 0$ of arbitrary phase space resolution by a single
quantum state is approached as the particle number $k$ tends to infinity.

\begin{figure}[htb!]
\begin{center}
\includegraphics[width=8.cm]{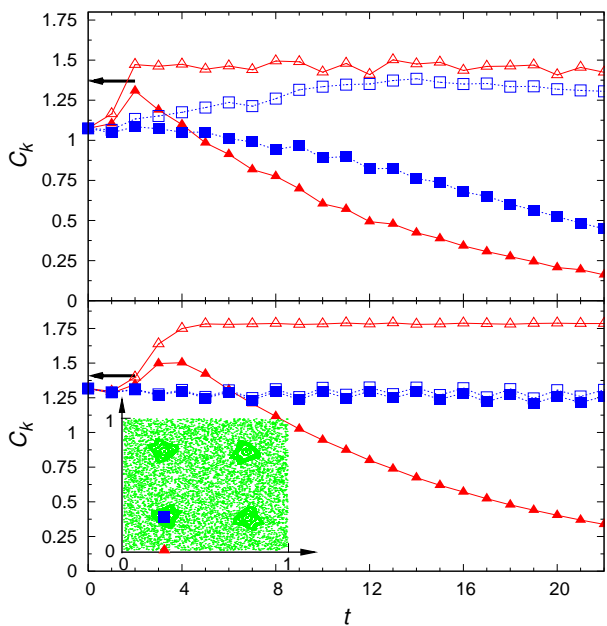} 
\end{center}
\caption{(Color online) Evolution of multipartite concurrence $C_k$ for two
different numbers of qubits: $k=5$ (top), $k=8$ (bottom).
Open symbols refer to unitary dynamics, while filled symbols represent
the evolution under the 
diffusive noise described by \equa{eq:dif}. Squares correspond to an
initial condition inside the  
nonlinear resonance island, triangles to initial conditions within the chaotic
sea, in the classical phase space spanned by $p$ and $q$ (see inset).  
For 
different qubit numbers, the ratio of noise strength to the effective
size of Planck's quantum, $\epsilon/\hbar_{\rm eff}$, 
is kept constant ($\epsilon=0.04$ for
$k=5$, and $\epsilon=0.005$ for $k=8$). 
%Thus, we used $\epsilon=0.04$ for
%$k=5$ and $\epsilon=0.005$ for $k=8$. 
Black arrows indicate the value of $C_k$ for $k$-partite GHZ states.
\label{fig:har}}
\end{figure}

Given this quantum coarse graining of phase space, we will now monitor the
time evolution of the multipartite entanglement of a quantum state initially
prepared as a 
minimum uncertainty Gaussian wave packet, launched at different positions on
the torus (which we will unfold as a unit square, for the ease of
illustration). We propagate the wave packet by the
unitary operator
\begin{equation}
U^t=(e^{i N \chi_2\cos(2\pi \hat{q})}e^{i N \chi_1\cos(2\pi\hat{p})})^t\, ,
\label{prop}
\end{equation}
(with $t$ integer)
which follows from the quantization of the classical Harper map
\begin{equation}
\label{eq:harper}
\begin{array}{rl}
p'=&p- \chi_1 \sin(2 \pi q)\\
q'=&q+ \chi_2 \sin(2 \pi p')
\end{array}\ \ {\rm (mod 1)}\, .
\end{equation}
Equation (\ref{eq:harper}) describes a particle subject to a periodic impulse, with a
position-dependent amplitude \cite{leboeuf}, and exhibits a transition from
regular to chaotic dynamics as the kicking strengths $\chi_1,\chi_2$ are tuned
from $\chi_1,\chi_2\lesssim 0.11$ to $\chi_1,\chi_2\gtrsim 0.63$. Here, we
employ 
the fixed values $\chi_1=\chi_2=\chi=0.4964$, what defines a mixed phase space
structure as depicted in the inset of Fig.~\ref{fig:har}. 
%%***********************************************************************
%%*********************************************************************** 

The quantum evolution (\ref{prop}) modifies the decomposition of the evolved
state $\ket{\psi_t}$ in the computational basis, and thus the entanglement
of the $k$ degrees of freedom defined by the constituent qubits. In order to
assess the nonclassical correlations inscribed into the evolved state, we use
the $k$-partite concurrence $C_k$ as defined in \cite{MCKB}. On pure states,
this quantity is given by the square root of a balanced average over the
squared concurrence of all nontrivial bipartitions of the $k$-set under
scrutiny, and vanishes exclusively for $k$-separable states
\cite{MCKB,rafal-2006}. 
Moreover, it has the particularly advantageous property  
$C_k(\ket{\phi}\otimes\ket{\psi}_{k-1})  =  C_{k-1}(\ket{\psi}_{k-1})$,
$\forall \ket{\phi}\in {\cal H}_1, \ket{\psi}_{k-1}\in 
{\cal H}_2\otimes\ldots\otimes{\cal H}_k$.
The latter allows to compare the entanglement inscribed into quantum states
composed of an increasing number of subsystems. Furthermore, $C_k$ has a
generalization for mixed states (through the convex roof construction
\cite{MCKB}), which we will make use of below.

We start out with a short inspection of the entanglement dynamics under purely
coherent dynamics, to set the scene. The evolution is generated by application
of $U^t$, where the integer $t$ 
counts the number of applications and defines a discrete time. The open symbols
in Fig.~\ref{fig:har} represent $C_k(t)$, for two different initial positions
(also indicated by filled
triangles and squares in the inset) of
the initial minimum uncertainty state in phase space  -- either within an
elliptic island ($q=p=0.25$; open squares), or within the chaotic phase space
component ($q=0.25$, $p=0.0$; open triangles). Since the computational basis
is encoded in position states, both initial conditions define the same initial
value
$C_k(0)$ ($C_5(0)=1.074386 $ and $C_8(0)=1.316826$, respectively)
of concurrence. Since, on the quantum level, chaotic dynamics is
tantamount of strong 
coupling in any basis, it immediately follows that $C_k$ will increase rapidly
for the initial condition placed in the chaotic domain, and saturate
once
equilibrated over the chaotic eigenstates
\cite{nemes1998,miller1999,fujisaki2003} of the quantized Harper map  
-- this is indeed observed in the figure. 
Note, however, that the saturation level does {\em not}
coincide with the maximal possible value of $C_k$ on the pure states, but
rather with its most probable value (which approaches the maximal value in the
limit $k\rightarrow\infty$ \cite{carlos}). Also note that, for $k>3$, this
value is {\em larger} than the $k$-partite entanglement of GHZ states,
which are maximally entangled only in the special case $k=3$ \footnote{This
 corrects an earlier statement in \cite{arrc}.}. Indeed, 
$C_{k=5}^{\rm GHZ}=1.369286$, and 
$C_{k=8}^{\rm GHZ}=1.408798$ (indicated by black arrows in
Fig.~\ref{fig:har}), approaching 
our minimal 
uncertainty state's initial multipartite entanglement with increasing $k$. Finally, the
qualitative behaviour of $C_k(t)$ for an initial state lying in the chaotic 
sea depends only weakly on the number of qubits,
as evident from a comparison of both panels in Fig.~\ref{fig:har}: For $k=8$
(corresponding to a Hilbert space dimension $N=256$), the time evolution is
very smooth, while for $k=5$ (Hilbert space dimension $N=32$) still some
fluctuations -- essentially a finite size effect -- are observed.

For the state initially placed within the elliptic island, size {\em
does} matter: for $k=5$, the initial coherent state cannot be
well accomodated within the elliptic island in phase space (due to the
finite size of $\hbar_{\rm eff}$), and exhibits non-negligible tunneling
coupling to its chaotic environment. Consequently, as time proceeds, the
coherently evolved state spreads more and more over the chaotic phase space
component, and its entanglement finally reaches essentially the same
value as for the initial condition within the chaotic domain, just after
considerably longer time -- essentially determined by the relevant tunneling matrix
elements (which, in general, will be strongly fluctuating under small
parameter changes \cite{ullmo,schlagheck}). In contrast,
for $k=8$, tunneling from the island into the chaotic sea occurs on a much
longer time scale (which, on average, increases exponentially with
$\hbar_{\rm eff}$), and remains invisible on the time scale covered in
Fig.~\ref{fig:har}. The small oscillations of $C_{k=8}(t)$ are due to the
spreading of the initial wave packet along the regular island's tori.

This screening of the initial state from the chaotic sea when initially placed
within the elliptic island, more and more efficient with increasing particle
number, has an immediate consequence for the robustness of the state's
multipartite entanglement under the influence of decoherence, as illustrated
by the filled symbols in Fig.~\ref{fig:har}: For chaotic initial conditions,
an initial rise of $C_k$ is rather quickly overruled by the loss of
multiparticle coherence and hence of entanglement, and this is once again
largely independent of $k$. However, for initial conditions within the island,
$k=5$ again leads to asymptotically the same behaviour as for the chaotic
initial condition, while $k=8$ induces entanglement dynamics almost completely
unaffected by the noise. Thus, for sufficiently large $k$, equivalent to
sufficiently small $\hbar_{\rm eff}$, and correspondingly suppressed tunneling
rates, the classical nonlinear resonance creates strongly entangled
multipartite states which, in addition, are robust against noise. This is
further illustrated in Fig.~\ref{fig:caca}, where $C_{k=8}(t=16)$ is plotted for
different initial conditions, in the absence and in the presence of noise:
%%***********************************************************************
\begin{figure}[htb!]
\begin{center}
\includegraphics[width=8.cm]{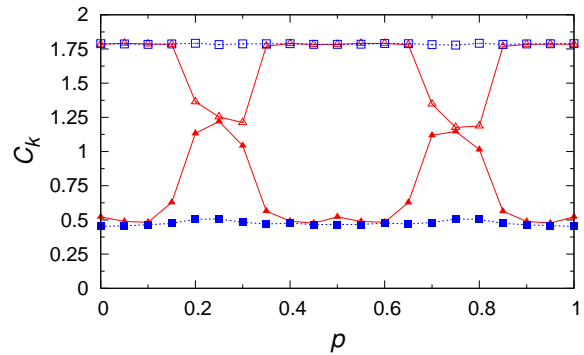}		%%{Ck-P.eps}  %  
\end{center}
\caption{(Color online)
Eight-partite concurrence $C_{k=8}$  as a function of initial momentum
($p=0.0, 0.05,\ldots,0.95,1.0$), after $16$ iterations of the Harper
map, with $\chi=0.4964$.  
Open symbols correspond to unitary evolution, while filled symbols
refer to unitary evolution amended by diffusive noise, 
\equa{eq:dif}. Squares represent the initial position $q=0.5$, and
triangles $q=0.25$.  
Peaks and dips of $C_{k=8}(t)$ are located exactly at the center of
the nonlinear resonance island in classical phase space.
\label{fig:caca}}
\end{figure}
%%***********************************************************************
Clearly, entanglement is robust when shielded by the resonance island. While
chaotic dynamics produce slighty stronger entanglement, this is significantly
more fragile under decoherence.

Given the above, some remarks on the applied noise are in order, since any
statement on the robustness of some kind of coherence {\em must} depend on the
nature of the noise. Here, we used Gaussian noise as described in
\cite{nonn,garma2}, which can be implemented experimentally by an enlargement
of the quantum register by a set of suitably initialized ancilla qubits
\cite{igm-unpublished}. 
It can be written as a map
\begin{equation}
\label{eq:dif}
\se(\rho)=\sum_{q p}\cpq \Tqp \rho\Tqpd\, ,
\end{equation}
which is applied once after each application of $U^1$. 
$\cpq$ is the discrete Fourier transform of
$\widetilde{c}_\epsilon(\mu,\nu)=\exp\left[
-\frac{1}{2}\left(\frac{\epsilon N}{\pi}\right)^2 
(\sin^2[\pi\mu/N]+\sin^2[\pi\nu/N])
\right]$
and is very close
to a periodic Gaussian of width $\epsilon/(2\pi)$, 
centered around $(q,p)=(0,0)$. 
The $\Tqp$ are unitary translation
operators on the torus \cite{schwinger}. 

The action of $\se(\rho)$ 
is easily understood in phase space: with high
probability, the state is left untouched, while with weight $\cpq$, {\em
  locally} in phase space,  every
possible translation is generated.  
This noise is similar to a high temperature bath of oscillators producing both 
diffusion and decoherence, and steeres the quantum dynamics into the
semiclassical limit \cite{andre2004}. 
In Wigner phase space representation, diffusion induces 
broadening and blurring of the 
contour of the state, while 
decoherence wipes out the interference fringes, and eventually transforms the
state into a mixed (classical) state. 
Given the $k$-dependence of
$\hbar_{\rm eff}$, the strength $\epsilon$ of the noise was scaled such as to
keep $\epsilon/\hbar_{\rm eff}$ constant, in the above plots. The observed
robustness of entanglement stems from the local action of  $\se(\rho)$, which
respects the classical phase space structure 
and leaves the initial state within the 
elliptic island effectively as a fixed point of the  
evolution, in the limit of large $k$ (small 
$\hbar_{\rm eff}$). The dramatically different dynamics of 
$C_{k=8}$ within and outside the island is a (multipartite)  
manifestation of enhanced decoherence in classically chaotic as 
opposed to regular systems \cite{zurek1994}, {\em for this specific type of 
noise\/}, and suggests the emergence of a 
subspace which is shielded against disentanglement, in the 
semiclassical limit.   

\begin{figure}[htb!]
\begin{center}
\includegraphics*[width=8.cm]{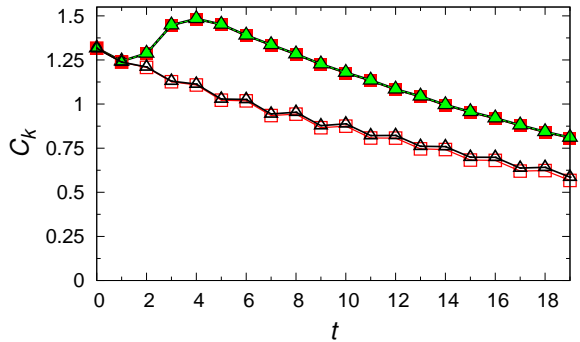} %%
\end{center}
\caption{(Color online) Evolution of multipartite concurrence
  $C_{k=8}$ 
%for $k=7$ 
in the presence of phase damping (triangles, (Eq.~(\ref{eq:pdc})) and depolarizing noise
(squares, Eq.~(\ref{eq:dpc})), with $\epsilon=0.04$ in Eqs.~(\ref{eq:pdc}) and
(\ref{eq:dpc}). Filled and open
symbols refer to initial conditions in the chaotic domain and within
the resonance island, respectively, precisely as in Fig.~\ref{fig:har}. 
\label{fig:pdc}}
\end{figure}

In contrast, we may choose {\em non-local} noise sources such as defined by the
multipartite phase
damping channel (PDC), 
\begin{equation}
	\label{eq:pdc}
\s^{PDC}_\epsilon(\rho)=(1-\epsilon)\rho+\epsilon\sum_i\rho_{ii}\ket{i}\brac{i}\, ,
\end{equation}
or the generalized depolarizing channel (DPC)
\begin{equation}
	\label{eq:dpc}
\s^{DPC}_\epsilon(\rho)=(1-\epsilon)\rho+\epsilon I\, .
\end{equation}
For a computational basis 
identified with position states, both these maps
can be expressed in terms
of translations on the torus \cite{colo}, though now with equal (rather
than Gaussian, see (\ref{eq:dif})) weight. Thus, they completely obliterate the
classical phase space structure, and couple different eigenstates of the
quantized Harper map, irrespectively of their localization properties in
regular or chaotic phase space domains. Accordingly, given the same initial
conditions as in Figs.~\ref{fig:har} and \ref{fig:caca}, concurrence decreases
monotonously with time, independently of the initial condition, except for the
typical short time transient (see also Fig.~\ref{fig:har}) observed for the
chaotic initial condition -- see Fig.~\ref{fig:pdc}. 
%%***********************************************************************
%%***************************************************************************************

To conclude, 
we have shown that minimum uncertainty states induce multipartite
entanglement in the associated computational basis, 
robust against the action of diffusive Gaussian noise, 
when launched within a nonlinear resonance island.
They can be produced
efficiently as ground states of the Harper Hamiltonian \cite{ronc},
subsequently 
translated using torus translation operators. 
The latter 
generate modular additions with 
controlled phase shifts on the register qubits, and can be implemented, e.g. 
in ion trap experiments
\cite{vedral,miqueladd}.
Furthermore, given the robust entanglement
evolution for initial conditions within a regular island, also those 
eigenstates of the Harper Hamiltonian 
which are anchored to the classical regular island exhibit the same
robustness properties. 
Thus, robust multipartite entangled states can
be defined through the resonance condition which defines the regular
island, a ubiquitous feature of
Hamiltonian systems with mixed classical phase space
structure.  

I. G.-M. was partially supported by CONICET (Argentina) and by the 
EC~IST-FET project EuroSQIP. F. M. acknowledges financial support of Alexander 
von~Humboldt foundation.
%%*************************************************************
%\bibliography{refs}
%%%%%%%%%%%%%%%%%%%%%%%%%%%%%%%%%%%%%%%%%%%%%%%%%%%%%%%%%%%%%%%%%%%%%%%%%%

%%%
\end{document}